# Scalable Organic Semiconductor Neutron Detectors

## Authors
Joanna Borowiec[1,2], Fani E. Taifakou[1], Muhammad Ali[1,3], Chris Allwork[1,3], Adrian J. Bevan[1,4](*), Theo Kreouzis[1], Cozmin Timis[1]

[1] Department of Physics and Astronomy, Queen Mary University of London, London, E1 4NS, United Kingdom

[2] College of Physics, Sichuan University, Chengdu 610064, China

[3] Atomic Weapons Establishment, Aldermaston, United Kingdom

[4] Alan Turing Institute, London, United Kingdom

(*) Corresponding Author e-mail: a.j.bevan@qmul.ac.uk

## Abstract
A long-standing limitation of semiconductor neutron detectors is the lack of a scalable solution to make large area instruments.  Neutron detectors are used in a wide range of applications, including the nuclear industry, safeguarding radioactive material, neutron imaging, non-destructive testing, understanding space weather effects on commercial electronics used in aviation, and for fundamental science such as nuclear and particle physics.  Here we demonstrate that a solution processed organic semiconductor technology, which is scalable, can solve this aproblem.  We have demonstrated detection of fast neutrons from monoenergetic beams (energies 0.565 and 16.5 MeV), from AmBe sources, and thermal energy (0.025 eV) in devices sensitised with $^{10}$B enriched $B_4C$. This energy range is of interest for many of the applications listed above. The detector response is found to be linear with flux up to $1.5 \times 10^7$ n/cm$^2$/s. As organic semiconductors are a similar density to human tissue, this technology may be of interest for medical physics applications, for example B neutron capture therapy dose monitoring and understanding neutron interactions in human tissue.

## Introduction
In the past few decades organic electronics became established as a commercial technology. Organic semiconductors (OSCs) are highly tuneable, easily processable and can be used to form printed, large-area, flexible, low-cost electronic devices. Applications for organic electronics have been found in a variety of areas; from organic light emitting diodes (OLEDs), organic photovoltaics (OPVs) and organic field effect transistors (OFETs) to bioelectronics. A novel application for organic electronics is as a radiation detector. While research in this field is limited, the ability to detect radiation using OSC systems has been demonstrated for α[1,2], β[3] and X-ray[4] radiation. Proton[5], fast[6] and thermal[7] neutron detection has also been reported. OSCs are low density, and consequently detection of highly ionising particles is efficient, whereas detection of β and X-ray radiation has lower efficiency.  The detection of fast neutrons is possible because of the high H and C content of OSCs. Sensitivity to thermal neutrons can be achieved by inclusion of isotopes such as $^{10}$B in the OSC or device, providing an alternative to $^3$He detectors, the cost of which is driven by limited supply and increasing demand[8, 9]. Another advantage of this technology is that OSC sensors can be operated efficiently at low voltage. We demonstrate the successful detection of both fast and thermal neutrons using OSC detectors using bias voltages from 1 to 100V.

Possible uses include: the nuclear industry where monitoring neutron fluxes in nuclear reactors (fusion and fission) is important to determine the operating condition of those reactors; nuclear security, where safeguarding radioactive material is ciricial to determing that waste is correctly stored or that a port of entry is able to detect trafficking of illicit material. Neutrons are also used in imaging and non-destructive testing, where large area segmented sensor arrays are used. Monitoring of space weather effects on commercial electronics used in aviation is important to ensure equipment is retired before end of life failure of the electronics from radiation damage occurs. Particle and Nuclear physics experiements use neutron detectors to measure the properties of the neutron or for example to detect the presence of neutron background for low background searches of dark matter or studies of the properties of neutrinos. Finally, as OSCs are the same density and of similar composition to living tissue, this technology may be of interest for medical physics applications such as B neutron capture (BNC) therapy dose monitoring and understanding neutron interactions in human tissue.

## Organic semiconductor detectors

The OSC neutron detectors used for this work are based on poly{[N,N'-bis(2-hexyldodecyl)naphthalene-1,4,5,8-bis(dicarboximide)-2,6-diyl]-alt-5,5'-(2,2'-bithiophene)} (PNDI). This is a high mobility OSC, often used as an electron acceptor in Donor-Acceptor devices[10]. We have tested devices fabricated on soda-lime glass as well as high- and low-density polyethylene substrates (illustrated in fig. 1) using the methodology presented in the supplementary information (SI). Glass substrate-based devices have been tested using indium-tin-oxide (ITO) anodes and aluminium cathodes. All substrates have been tested using aluminium for both electrodes. Two types of devices were tested, type 1 (without) and type 2 (with) $B_4C$ sensitisers dispersed in the PNDI. The device thickness varied between ~5 and ~50 µm. Prior to neutron exposure, devices are characterised by performing current-voltage measurements. Furthermore to assess the signal to noise ratio (SNR) of devices, these were exposed to α particles from an $^{241}$Am source. Devices with the highest SNRs were selected to be tested with neutrons. In contrast to our previous work[2], which reported a maximum SNR of 32:1 on exposure to $^{241}$Am α particles, these devices achieve values up to 3000:1. The corresponding dark currents are as low as ~10 pA for bias voltages of ±100 V, compared with ~1nA for our previous work, and additionally are stable after a short conditioning period under bias. A significant increase in current emitted from the device when exposed to radiation can be classified as a detection event, the current increase being the signal.

## Neutron sources and interactions

Detection of neutrons is more complicated than for other types of radiation as the detection method depends on the isotope content in the sensor and incident neutron energy (see SI). Thermal neutrons have energies of ~0.025 eV, epithermal neutrons used in medical applications have energies of ~0.025-0.4 eV, and fast neutrons have energies above ~1 MeV. Detection of fast neutrons via elastic scattering in OSCs is dominated by proton recoil from scattering of neutrons with H nuclei at low energies.

As the neutron energy increases, there are additional contributions from C nucleus recoil and nuclear interactions, as illustrated in fig. 1.

For thermal neutron detection the elastic scattering is non-ionising as there is insufficient energy in the collision to create electron-hole pairs. Thus, one must add isotopes with enhanced thermal neutron capture cross sections and rely on detecting the capture decay products. Our detectors are based on an OSC (the polymer PNDI) component with and without the inclusion of isotope enriched $B_4C$ microparticles. The isotope $^{10}B$ has a thermal neutron capture cross section of 3840 barns, and decays into an α particle, a $^7Li$ nucleus. We use $B_4C$ with 96.6% enrichment of $^{10}B$ to detect thermal neutrons, compared to 20% for naturally occurring B.

Where the neutron field contains a broad energy spectrum, our detectors will respond to different energy neutrons by different processes as noted above. By studying the difference in signals between devices with and without $B_4C$ we can differentiate between the fast and thermal neutron device response.

We use the GEANT4 Monte Carlo simulation framework[11] to understand the interaction of fast and thermal neutrons in OSCs. The high precision neutron interaction library is used to simulate neutron interactions for each of the isotopic components in our devices[12, 13, 14]. Details of the simulation can be found in the SI. This tool is used to evaluate potential contributions (i.e. energy loss in the device) from neutron and secondary interactions with different isotopes in the detector material and the housing containing the detector (see SI fig. 6). We can additionally understand the fast and thermal neutron responses resulting from different types of hadron interactions (proton i.e. $^1H$, α, $^7Li$, and $^{12}C$) and how they contribute to the signal. The ratio of average $^1H$ to $^{12}C$ recoil energy loss in the detector calculated for 0.565 and 16.5 MeV neutrons is found to be 9 and 0.5, respectively. This indicates that lower energy neutrons interact more readily with $^1H$ than C in the polymer. Above 6 MeV hadronic channels open resulting in $^{12}C$ decay, which also contributes to the energy deposited in the device.

### Experimental work

In this article we present results of exposing OSC detectors to thermal and fast neutron beams at the National Physical Laboratory (NPL) in the UK[15, 16]. The fast neutron beam energies used are 0.565 and 16.5 MeV with neutron fluxes of 1.6 – 2.0×10⁶ and 2.3 – 2.8×10⁵ n/cm²/s, respectively. Devices are exposed to thermal neutrons in the access hole at the NPL thermal pile, with neutron fluxes up to 1.3×10⁷ n/cm²/s. We note that signals from the thermal pile have a significant fast neutron component. We use devices with and without $B_4C$, with different substrates as well as cadmium shields around individual devices to disentangle the fast neutron signal from any thermal or epithermal neutron contribution for our measurements as described below.

### Fast neutron detection

We have studied the fast neutron detection performance of type 1 devices (PNDI with no $B_4C$ sensitiser) using 0.565 MeV and 16.5 MeV fast neutron beams. The results obtained from these exposures are shown in fig. 2 panels a and b respectively. The current through the device is monitored continuously and as the neutron flux is alternately switched on and off a clear response to the presence of neutrons is evident. The magnitude of the current is increased in the presence of neutrons (the beam is on between 30 and 50, 70 and 90 s etc.).

The plots in figs 2a and b also demonstrate the device bias dependence of the response (at -30 V and -100 V). The device performance is the same in forward and reverse bias. Variations in the device current during individual neutron exposures (e.g. between 30 and 50 s in fig. 2a -30 V, or between 30 and 50 s fig. 2b for both biases) are due to real variations in the neutron flux as a result of relatively slow beam feedback control. This has been confirmed using the NPL beam monitoring instruments (also see SI, fig. S9). We detect fast neutrons for both beam energies tested and we conclude that we observe fast neutrons arising predominately from elastic scattering off $^1$H for 0.565 MeV. For 16.5 MeV neutrons the elastic scatter contribution is dominated by $^{12}$C recoil. The magnitude of signal is similar for both energies, whereas the 0.565 MeV beam is an order of magnitude higher flux than the 16.5 MeV beam. This indicates that the OSC device is sensitive to neutron energy. We have additionally observed the response to high energy neutrons (up to ~11 MeV[17]) using AmBe sources (see SI fig. S8) which does not suffer from the aforementioned beam artifacts.

Modelling suggests that a high density polyethelyne (HDPE) substrate should enhance the energy deposited within the semiconductor, compared to a glass one for fast neutrons (see SI, fig. S5). However no significant difference was observed for the devices tested. Full understanding of the response to neutrons above 6 MeV requires further study given the multiple elastic and inelastic processes involved. Devices fabricated with different electrode combinations (aluminium or ITO) yield similar results, eliminating the possibility of significant contributions to the signal arising from fast neutron resonances with higher atomic number constituents in the electrodes. To eliminate any possible thermal neutron contamination, we used a cadmium shield to block neutrons with an energy less than ~0.5 eV. The fact that a signal remains with shielding in place demonstrates that these devices are indeed responding to fast neutrons.

### Thermal neutron detection

Figure 3a shows the response for neutrons detected in the access hole of the thermal pile for type 1 and type 2 devices at -100 V bias. A clear signal enhancement is observed for the type 2 device compared to the type 1 device. While some device-to-device variation is expected, it is negligible compared to the signal difference obtained. Type 1 devices are only sensitive to fast neutrons (via elastic and inelastic scatter), whereas the type 2 devices (which contain $^{10}$B enriched $B_4C$) will detect both fast neutrons as well as thermal neutrons via the BNC process. From fig. 3a it is clear that $^{10}$B enhances the response in type 2 detectors, more than doubling the signal obtained with a type 1 detector (both at the same bias).

Figure 3b shows the response of the same type 2 device as shown in 3a, to variations in the average neutron flux, ranging from $1.3 \times 10^6$ to $1.3 \times 10^7$ n/cm$^2$/s. The linearity between the detector response and the flux has been confirmed (see SI, fig. S7). A neutron beam flux overshoot is seen for the thermal pile.

Figure 4a shows the thermal pile type 2 device current enhancement versus bias voltage. The response increases with bias, indicating enhanced charge extraction. Figure 4b shows the SNR obtained for a range of bias voltage up to ±100 V for a fixed neutron flux of $1.3 \times 10^7$ n/cm$^2$/s. The SNR remains above 50 for all bias voltages used with these devices, reaching 750 for +100V.

To rule out substrate contributions to the signal, for example B content of the glass or polymer substrate enhancement of the signal, we study devices fabricated on HDPE compared with glass, and also (see SI) perform material analysis of the stubstrates. Figure 4c shows the thermal pile responses of two type 1 devices: one using a glass substrate and the other HDPE. The responses measured under identical conditions (-100 V bias and the same beam flux) differ by 1.4 standard deviations (as calculated using the spread of the responses to individual neutron exposure periods), indicating that there is no significant difference between the response obtained using the glass substrate from that obtained using the HDPE. We conclude that the type 1 detector response obtained in the thermal pile (fig. 3a) is not significantly affected by the choice of glass or HDPE substrate.

Using the $^{10}$B thermal neutron cross section, the $^{10}$B average atomic number density and the device thickness we can calculate the probability of thermal neutron capture in the type 2 device as (18±3) %. This value provides an upper limit on any detection efficiency and is referred to by us and other authors as a "quantum efficiency"[7]. We use the difference in signal current between the two types of devices to estimate a thermal neutron conversion efficiency of 2%, physically acceptable but significantly lower than the quantum efficiency. The difference can be explained by clumping of the $B_4C$ as shown in SI fig 11, which may prevent BNC decay products created deep within a cluster from reaching the OSC.

If we used B with natural isotopic abundance instead of the enriched material, we would expect our signal to be dominated by the fast neutron response arising from proton recoil off the $^1$H in the OSC due to the high energy neutron tail in the beam produced in the thermal pile. We note that in previous published work by P. Chatzipiroglou et al.[7] using NPL's thermal pile with devices including a natural isotopic abundance B sensitiser did not account for the fast neutron OSC response.

## Conclusions

We have demonstrated that OSC detectors respond to monochromatic fast neutrons with energies 0.565 and 16.5 MeV, as well as neutrons from an AmBe source. Using the thermal pile at NPL we have shown that deploying OSC detectors both with and without $B_4C$ sensitisers can enable the determination of thermal and fast neutron components of the beam. We find that the signal magnitude scales linearly to a thermal neutron flux of $1.3 \times 10^7$ n/cm$^2$/s.

This set of measurements highlights the potential for OSC technology to detect neutrons over 8 orders of magnitude in energy from 0.025 eV through to 16.5 MeV. This is of interest for a range of applications discussed in the introduction. The different signal currents normalised to flux for fast and thermal neutron components in the thermal pile and the fast neutron beams suggest an energy dependent response of the detector technology that merits further study. Results presented are for high flux responses of our detectors resulting in a current increase. As demonstrated in F. E. Taifakou[2], even higher sensitivity can be achieved by, for example, averaging repeated measurements to minimise noise. The detectors presented here use the same fabrication process used in commercially available large area organic electronic systems, demonstrating a semiconductor neutron detector technology that overcomes the scalability limitation of existing technology.

## Acknowledgements
The authors gratefully acknowledge the generous contributions of the following UK organisations, AWE Plc., Queen Mary University of London, and the Science and Technology Facilities Council (grant numbers ST/S00095X/1, ST/T002212/1, ST/V000039/1 and ST/T002212/1) to support this work. Prof. Bevan was also supported by The Alan Turing Institute under the EPSRC grant EP/N510129/1. Furthermore we acknowledge the Technical Staff within the School of Physical and Chemical Sciences at Queen Mary University of London for technical support and assistance in completing this work. The authors thank P. Knight for insightful feedback on this paper and Dr. D. J. Thomas and his colleagues at NPL for providing detailed information about the experimental conditions available at the thermal pile that have enabled the analysis and interpretation of the data presented.

## References

[1] P. Beckerle and H. Ströblel, Charged particle detection in organic semiconductors, Nucl. Instrum. Meth. A **449** (2000) 302-310.

[2] F. E. Taifakou et al., Solution-Processed Donor–Acceptor Poly(3-hexylthiophene):Phenyl-C61-butyric Acid Methyl Ester Diodes for Low-Voltage α Particle Detection, ACS Appl. Mater. Interfaces, **13**, 5, (2021) 6470–6479.

[3] T. Suzuki et al., Organic semiconductors as real-time radiation detectors, Nucl. Instrum. Meth. A **763** (2014) 304-307.

[4] F. A. Boroumand et al., Direct X-ray Detection with Conjugated Polymer Devices, Appl. Phys. Lett. **91** (2017) 033509-033511.

[5] I. Fratelli et al., Direct detection of 5 MeV protons by flexible organic thin-film devices, Sci. Adv., **7** (16), (2021) eabf4462.

[6] A. Kargar et al., Organic Semiconductors for Fast-Neutron Detection, Proceedings of the IEEE Nuclear Science Symposium (2011) doi:10.1109/NSSMIC.2011.6154732.

[7] P. Chatzipiroglou, J. L. Keddie, and P. J. Sellin, Boron-Loaded Polymeric Sensor for the Direct Detection of Thermal Neutrons, ACS Applied Materials & Interfaces **12** (29) (2020) 33050-33057.

[8] R. T. Kouzes et al., Neutron detection alternatives to $^3$He for national security applications, Nucl. Instrum Meth. A **623** (2010) 1035-1045.

[9] R. T. Kouzes et al., Progress in alternative neutron detection to address the helium-3 shortage, Nucl. Instrum. Meth. A **784** (2015) 172-175.

[10] S. Kim et al., Rationally Designed Donor–Acceptor Random Copolymers with Optimized Complementary Light Absorption for Highly Efficient All-Polymer Solar Cells, Adv. Funct. Mater., **27**, 1703070 (2017); DOI: 10.1002/adfm.201703070.

[11] J. Allison et al., Recent developments in GEANT4, Nucl. Instrum. Meth. A **835** (2016) 186.

[12] E. Mendoza, D. Cano-Ott, T. Koi and C. Guerrero, New standard evaluated neutron cross section libraries for the GEANT4 code and first verification, IEEE Trans. Nucl. Science **61** (2014) 2357.

[13] E. Mendoza, D. Cano-Ott, C. Guerrero, and R. Capote, New Evaluated Neutron Cross Section Libraries for the Geant4 Code, IAEA technical report INDC(NDS)-0612, Vienna, 2012. Data available online at Geant4.

[14] E. Mendoza, D. Cano-Ott, Update of the Evaluated Neutron Cross Section Libraries for the Geant4 Code, IAEA technical report INDC(NDS)-0758, Vienna, 2018.

[15] P. Kolkowski and D. J. Thomas, Measurement of the fast neutron component in the beam of the NPL Thermal Neutron Column using a Bonner sphere spectrometer, NPL REPORT CIRM **28** (1999).

[16] N. P. Hawkes et al., Additional Characterisation of the thermal neutron pile at the National Physical Laboratory, UK, Radiation Protection Dosimetry Vol. **180**, No. 1–4 (2018) 25–28.

[17] J. W. Marsh, D. J. Thomas, M. Burke, High resolution measurements of neutron energy spectra from Am-Be and Am-B neutron sources, Nucl. Instrum. Meth. A **366** (1995) 340-348.

## Author contributions
Paper Writing: MA, CA, AB, JB, TK, FET, Simulation: MA, CT, FET Fabrication: JB, FET, Lab and Field Tests: AB, JB, TK, FET, CT, Analysis: AB, JB, TK, FET, Interpretation: AB, FET, TK, Project oversight: AB, TK

## Data access statement
All relevant data are provided with the paper and its SI.

## Figures

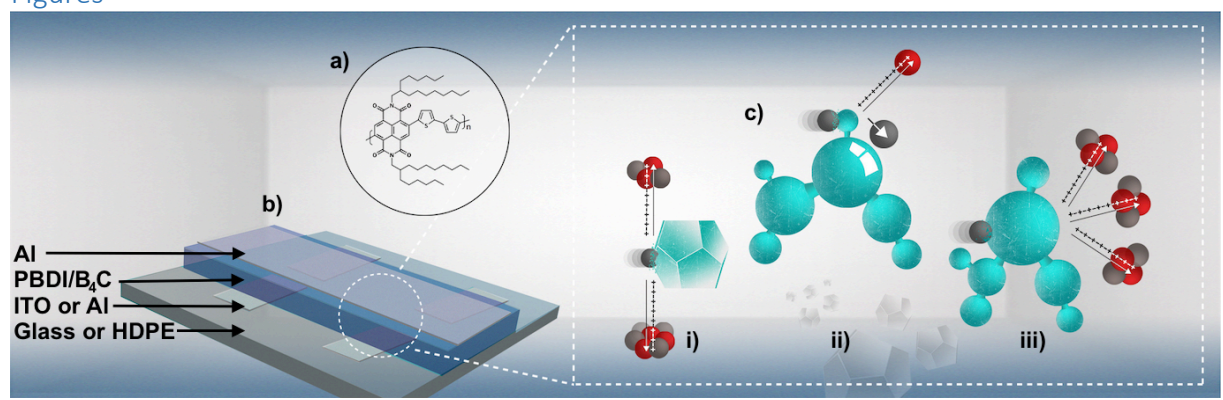

**Figure 1:** a) Chemical structure of the polymer OSC poly{[N,N'-bis(2-hexyldodecyl)naphthalene-1,4,5,8-bis(dicarboximide)-2,6-diyl]-alt-5,5'-(2,2'-bithiophene)} (PNDI). b) Schematic of four diode devices fabricated on a single substrate. The overlap areas between the anodes and the common cathode define individual devices. c) Schematic cross section through a single device. The current direction is from anode to cathode. Ionising (charged) neutron reaction products increase the charge carrier density within the device thus increasing the device current. Three schematic neutron interactions are indicated as i) thermal neutron $^{10}$B capture at a $B_4C$ microparticle, ii) elastic neutron-proton scattering at a PNDI $^1$H site, iii) $^{12}$C decay into 3 α particles at a PNDI C site.

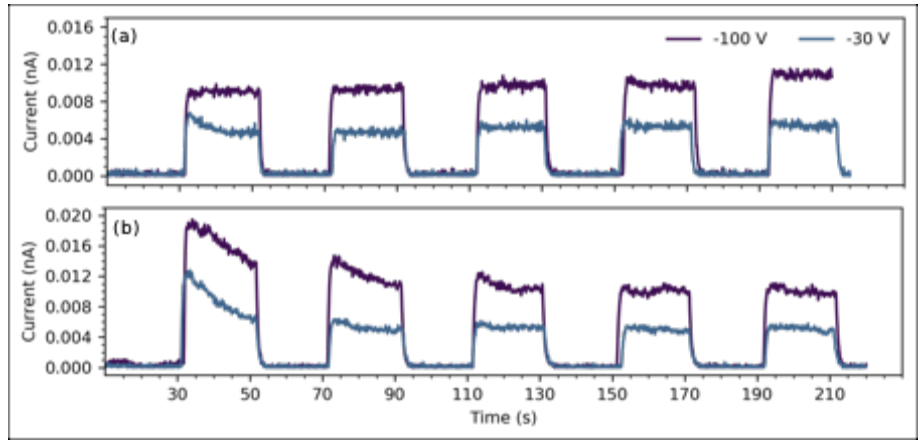

**Figure 2:** Device current magnitude versus time plots in the absence and presence of neutrons. The current enhancement due to the neutron flux is clearly distinguishable as a sharp rise on exposure, followed by a sharp drop when the neutron source is turned off. a) Type 1 device (HDPE/Al/PNDI/Al) under 565 keV neutron exposure. b) Type 1 device (HDPE/Al/PNDI/Al) under 16.5 MeV neutron exposure. For a) and b) the response at two different device biases is shown. All plots have been pedestal (background) corrected for clarity.

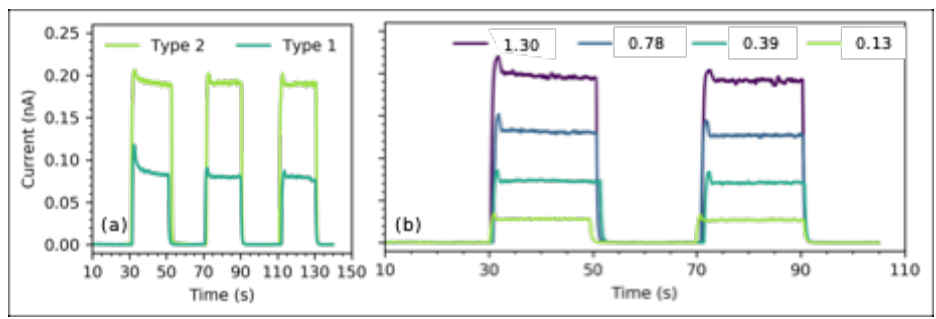

**Figure 3:** a) Thermal pile neutron exposure results of a type 2 (glass/ITO/PNDI:$B_4C$/Al) and a type 1 (glass/ITO/PNDI/Al) device, both at -100 V bias. b) The type 2 (glass/ITO/PNDI:$B_4C$/Al) device response at -100 V bias, parametric in neutron flux. The fluxes noted are in units of $10^7$ n/cm$^2$/s. All plots have been pedestal (background) corrected for clarity.

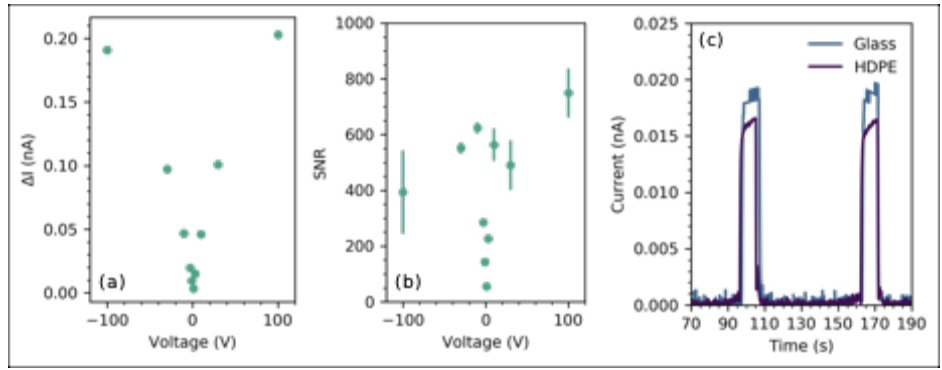

**Figure 4:** a) Thermal pile type 2 (glass/ITO/PNDI:B$_4$C/Al) device current enhancement (ΔI) versus bias voltage. b) Thermal pile type 2 device signal to noise (SNR) over the bias range of [-100, 100] V. c) Type 1 device current at 1.3 × 10$^7$ n/cm$^2$/s using different substrate materials (glass and HDPE). The responses agree to within 1.4 standard deviations.

# Supplementary Information: Scalable Semiconductor Neutron Detectors

## Authors
Joanna Borowiec[1,2], Fani E. Taifakou[1], Muhammad Ali[1,3], Chris Allwork[1,3], Adrian J. Bevan[1,4](*), Theo Kreouzis[1], Cozmin Timis[1]

[1] Department of Physics and Astronomy, Queen Mary University of London, London, E1 4NS, United Kingdom

[2] College of Physics, Sichuan University, Chengdu 610064, China

[3] Atomic Weapons Establishment, Aldermaston, United Kingdom

[4] Alan Turing Institute, London, United Kingdom

(*) Corresponding Author e-mail: a.j.bevan@qmul.ac.uk

## Methods

### Materials

poly{[N,N'-bis(2-hexyldodecyl)naphthalene-1,4,5,8-bis(dicarboximide)-2,6-diyl]-alt-5,5'-(2,2'-bithiophene)} (PNDI(2HD)2T), referred to as PNDI in the paper, is used as received from Ossila[1]. The HOMO and LUMO levels of -5.80 and -3.82 eV, respectively[2]. Figure S1 shows the molecular structure of PNDI.

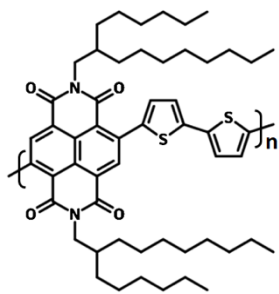

**Figure S1**: The molecular structure of PNDI(2HD)2T.

We use $B_4C$ sourced with a $^{10}B$ enrichment of 96.64% by atomic weight, a factor of 4.9 times larger than that found in the naturally occurring isotope. The particle size of the sensitisers that we mix into the OSC is <5 µm with this batch of material having an average particle size of 0.65 µm. The material purity is 99.924 %, and the largest contaminants are aluminium (380ppm), silicon (250 ppm) and iron (130 ppm). The motivation for using $^{10}B$ enriched material is the high cross section for thermal neutron interactions (3840 barns). The dominant isotope, $^{11}B$ has a negligible thermal neutron interaction cross section of 0.05 barns in comparison. The boron neutron capture process (BNC) proceeds via

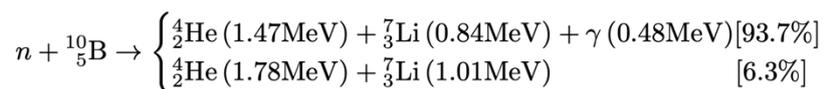

$$n + {}^{10}_{5}\text{B} \rightarrow \begin{cases} {}^{4}_{2}\text{He}\,(1.47\text{MeV}) + {}^{7}_{3}\text{Li}\,(0.84\text{MeV}) + \gamma\,(0.48\text{MeV}) & [93.7\%] \\ {}^{4}_{2}\text{He}\,(1.78\text{MeV}) + {}^{7}_{3}\text{Li}\,(1.01\text{MeV}) & [6.3\%] \end{cases}$$

The 0.48 MeV γ signal is neglected in our interpretation of device response as most photons escape the thin low mass detector without interacting.

**Elemental composition:** We have studied the elemental composition of the different materials and components in our devices using X-ray photon spectroscopy (XPS). This enabled us to verify that there was no cross-contamination of our type 1 devices with $B_4C$ or other sources of B. This check was essential as both types of devices are fabricated in the same glove box. This set of tests on all constituent materials verifies that the only boron containing material in our setup was the PNDI:$B_4C$ mix from our type 2 devices. The total absence of other elements that have isotopes with high thermal neutron cross sections (e.g. $^6$Li, $^{157}$Gd, and $^{113}$Cd) was also confirmed. Therefore, we conclude from the data shown in the main article that: given the large difference between type 1 and type 2 devices results from the $^{10}$B enhancement; type 1 devices are essentially blind to thermal neutrons, as expected, and therefore measure the fast neutron tail in the neutron beam. Example XPS spectra are shown in fig. S2 for PNDI, $B_4C$ and the glass substrate. The B1s peak corresponds to boron and is only evident in the $B_4C$ sample tested.

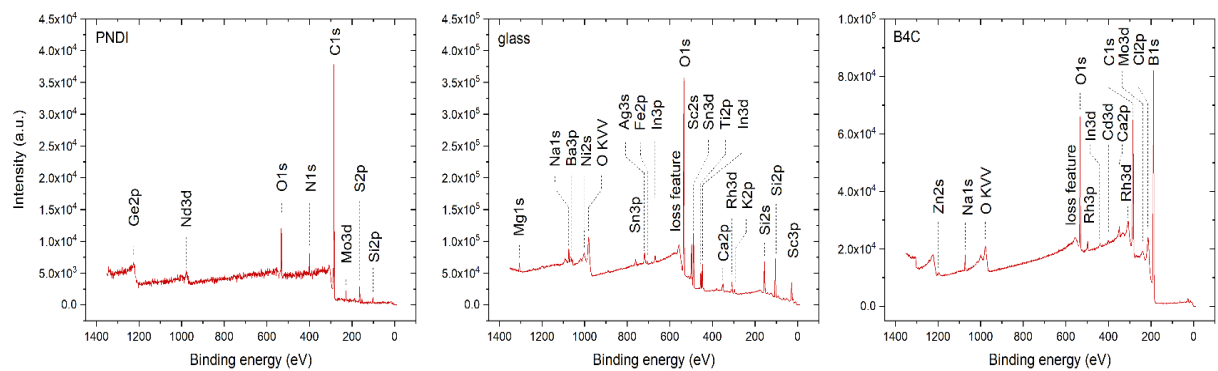

**Figure S2**: The XPS energy spectra determined for the PNDI (left), the glass substrate (middle) and the $B_4C$ (right) used.

Systematic effects relating to the signal dependence on the choice of electrode material, indium-tin-oxide (ITO) or aluminium (Al), and one the substrate material, glass, or high-density polyethylene (HDPE) have also been studied as discussed below.

Samples
Samples are made by first fabricating the bottom electrode (ITO or Al) on the substrate. For glass substrates this is done via photolithography to etch the ITO pattern. Where the bottom electrode is Al, this is deposited through a shadow mask using vacuum deposition. The organic semiconductor, or organic semiconductor-$^{10}B_4C$ components of our devices are dissolved in a 1:1 (volume) mixture of chlorobenzene and dichlorobenzene (typically 8 mg/mL) and stirred at 60 °C. In the case of $^{10}B_4C$ containing devices, the $^{10}B_4C$ is initially dispersed in the solvent using an ultrasonic probe. The solution is drop cast onto the substrate. This process is performed on a hotplate at 60°C to accelerate solvent evaporation, leaving a dry semiconductor film on the substrate. For the samples presented in this paper the drop casting process is repeated approximately 4 times to achieve the desired device thickness. Finally, an Al electrode is applied to the top surface of the device via vacuum deposition.

The thickness of device is tuned by polymer concentration in the solvent, and the number of times that the drop casting step is repeated. Here we report on devices that are between 35 and 50 μm thick. The $^{10}B_4C$ is included up to 20% by weight for the type 2 devices.

## Simulation

We use the GEANT 4 Monte Carlo simulation framework[3] to model our devices. The geometry model includes volumes corresponding to the substrate, electrodes, and the OSC material. The simulation relies on a probabilistic interaction model based on an accurate implementation of the isotopic content of the constituent materials assigned to a given volume of the object under study. Incoming particles are generated with known direction and energy and stepped through the detector model. At each step the Monte Carlo simulation computes the probability of a given set of interactions that include boron neutron capture (BNC) and elastic scatter processes (illustrated in fig. S3). The high precision neutron simulation library is used for our work[4,5,6].

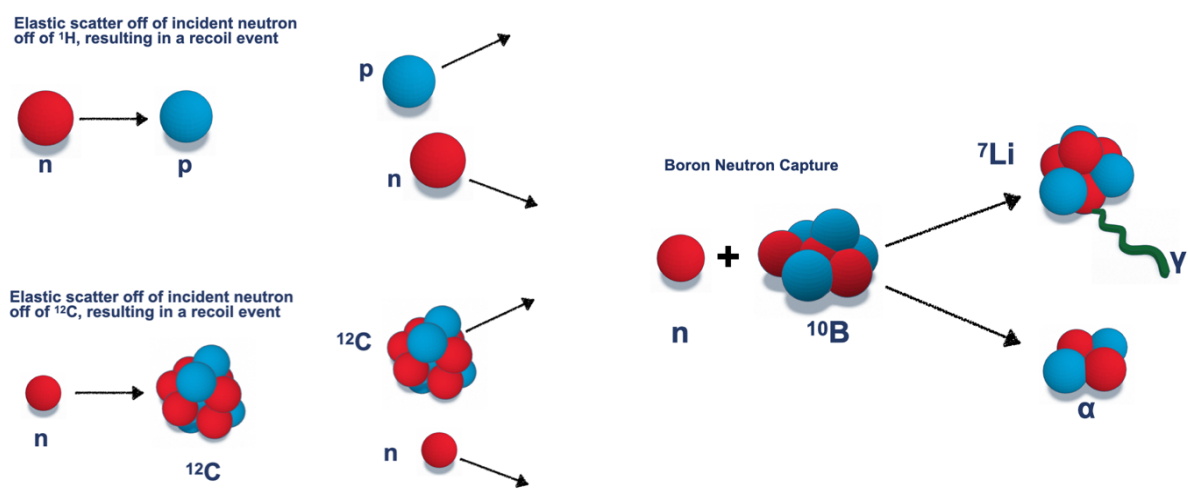

**Figure S3**: Illustrations of neutron elastic scatter off a (top left) proton, (bottom left) $^{12}C$ nucleus, and (right) the BNC process, showing the photon emission that occurs in 94% of capture events.

We use GEANT4 version 10.07, and track the energy deposited in the active region of our detector from different sources. Namely α and $^7Li$ from boron neutron capture (neglecting the energy deposited by photons as that is negligible given the thickness of the detector and density of the OSC) and proton and carbon nuclei recoil from elastic scattering. Inelastic scattering for the 16.5 MeV neutron beam is also simulated. The cross sections for elastic and inelastic scatter of neutrons with $^1H$ and $^{12}C$ up to 20 MeV are shown fig. S4.

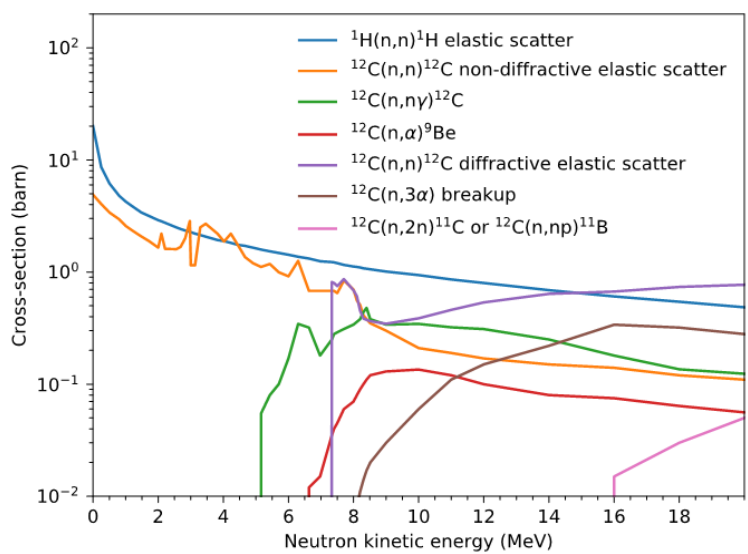

**Figure S4**: Cross sections of different elastic and inelastic neutron scatter processes. The dominant contribution at low energies comes from proton recoil ($^1$H elastic scatter). The relative importance of $^1$H to $^{12}$C elastic scatter depends strongly on energy. Inelastic $^{12}$C processes open above 6.5 MeV. The neutron interaction cross sections are taken from[7].

The effect of substrate material composition was tested by substituting the glass with HDPE substrates. The results of which are shown in fig. 4c in the paper. A GEANT4 study of the energy deposited in glass and HDPE from proton recoil ($^1$H elastic scatter) and heavy particles (namely $^{12}$C inelastic scatter) is shown in fig. S5 for both substrate types. The underlying energy spectrum assumed corresponds to that of fig. 3 of Ref[8], which is the measured energy spectrum for the thermal column. Note that this differs from the spectrum expected in the access hole used for our thermal neutron measurements. However, while expected to be similar, the precise energy spectrum of the access column is currently not well known. The heavy particle energy contributions are similar for both, and a small difference in the proton recoil energy deposit is evident, resulting in an expected 12% increase in energy deposited for devices built on a HDPE substrate compared to a glass one.

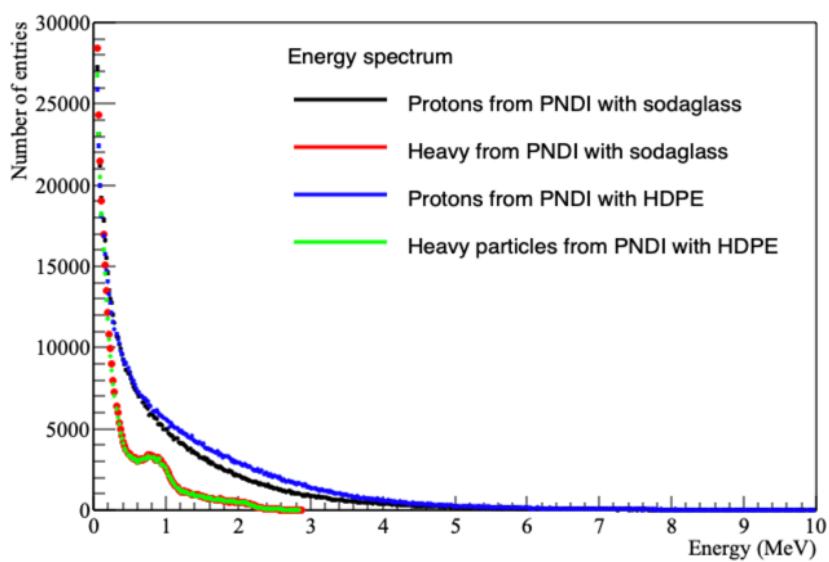

**Figure S5:** Histogram of energy deposited within the PNDI layer generated using a random selection of neutron energies consistent with the expected spectrum of fast neutrons in the thermal column at NPL, i.e. from the high energy "tail" of the thermal pile spectrum for different substrates (HDPE and glass) and by proton scattering and heavier nuclei scattering.

Measurement process

Measurements are made using a Keithley 2635B source measure unit, which has a 0.1 fA resolution for current measurements and can supply a bias of up to ± 200 V. Our detectors are biased and undergo an initial settling period. The reason for this is that semiconductor devices can have many charge carrier trap states, by applying a bias to the device, those trap sites are filled routinely (as evidenced by the initial current drift). Once the device response has settled, typically after a few tens of seconds, they are exposed to radiation. Results have been obtained for neutron detection at the National Physical Laboratory (NPL) in Teddington, UK, with bias voltages ranging from ± 1 V to ± 100 V. Higher bias voltages have been used when testing devices with $^{241}$Am α source in the laboratory. We measure one device at a time and compare between different exposures according to the beam flux delivered by the accelerator beamline used (either high energy or thermal).

Signal to noise determination

We use a Gaussian regression fit to the data when devices are not exposed to radiation to determine an "off" baseline. The baseline corrected data are subsequently analysed to determine the signal to noise ratio. For the PNDI results shown here there is very little device drift with time and one can observe signals clearly without the need for baseline correction. The baseline correction method also removes the residual dark current contribution, again small in PNDI (a few to tens of pA). The root-mean-square (RMS) of the beam on data from the Vann de Graff accelerator is dominated by variations in the beam current and affected by an initial current overshoot that is quickly compensated for higher demand level (flux) beam settings. The beam current variation dominates the flux, and hence signal distributions obtained. This is not a good measure of device performance, as evidenced by the source-based measurements that have a much smaller RMS when devices are exposed to radiation compared to those from the accelerator. The RMS of signal about the baseline (beam off) is used to derive the noise for neutron beam data to dis-entangle device performance from the beam current effect.

### Neutron beams and sources

The on/of exposure of devices to neutrons is controlled manually by virtue of (for beams) the accelerator operator withdrawing or inserting a beam stop in the Van De Graff accelerator to switch on or block the beam. When using the monochromatic neutron beam energies of 0.565 and 16.5 MeV, the detector was placed approximately 4.5 cm from the target, in a HDPE housing (to minimise material activation) covered with conductive paint, as shown in fig. S6. Neutrons with an energy of 0.565 MeV are produced using $^7$Li on a Be target, and 16.5 MeV neutrons are produced using a deuterium beam on a tritium target. The proximity of the detector to beam target results in an energy spread of the beam from on to off axis. Our devices are 2 mm square. The beam energy and resulting spread is dominated by the positioning uncertainties of the detector to the beam axis. Assuming a 10 mm positioning uncertainty, the expected energy range of neutrons exposed to the devices is 0.552 – 0.565 and 16.42 – 16.50 MeV.

Details of the neutron flux and systematic uncertainties are provided by the NPL facility for the experimental data collected. The neutron flux obtained is between 1.6 – 2.0×10$^6$ and 2.3 – 2.8×10$^5$ n/cm$^2$/s for the two energies, respectively, depending on measurement run. Systematic uncertainties on the flux from beam has been provided by the NPL facility. This is dominated by position uncertainty (between 9 and 10 %). An additional flux uncertainty of 9.5 % (0.565 MeV) and 10.0-10.7 % (16.5 MeV) results from several sources. This latter systematic uncertainty is determined using a calibration detector with contributions arising from variation in the neutron fluence with incident beam current, efficiency and positioning of the calibration detector, long term stability of that detector, and statical uncertainties in the flux determination. In addition, there is a 0.5 % uncertainty in the in-scatter correction when measuring flux for the 16.5 MeV beam. At 95 % confidence level the flux uncertainty is 19-21 %.

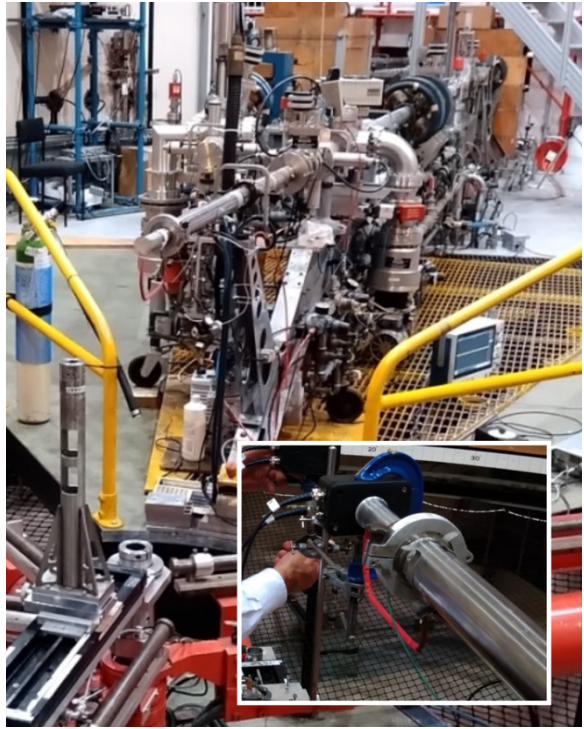

**Figure S6:** The NPL high energy neutron beam line, (inset) detector housing positioned next to the beam target at the end of the beam line.

Neutron interactions with materials such as OSC result in elastic and inelastic neutron scattering. A corollary of this is a softening of the neutron energy spectrum. Moderation of neutrons from monoenergetic beams can affect results even at a facility such as NPL. It is possible to determine the response of a detector to a monoenergetic beam by measuring the nominal signal and performing a second measurement using a shadow cone to determine the background. Subtracting the two (appropriately normalised) allows one to disentangle moderated neutrons from the fast neutron signal. It was not possible to use this approach for the measurements presented in this paper due to the flux response of the devices under test, however the fraction of scattered neutrons measured in the monoenergetic beam is known from other measurements at NPL. The scatter component of the monoenergetic neutrons beam is < 2 % for 0.565 MeV and < 5 % for 16.5 MeV. While this residual background is small, we further validated that our detectors respond to the fast neutron component of the beam by performing type 1 device measurements in fast neutron beams with and without a cadmium shield around the device holder. The shield absorbs all neutrons with energies below 0.5 eV, providing experimental verification that are detecting neutrons above this threshold.

The thermal pile at NPL provides a radiation field dominated by thermal neutrons. There is a residual tail of fast neutrons that is not well known for the access column but assumed to be similar to that found in the thermal column. The same features of beam operation that affect fast neutron measurements are manifest for the thermal neutrons (e.g. variance due to beam currents and an initial overshoot in flux). We accumulate data at different fluxes of 0.13, 0.38, 0.78, 1.3 ×$10^7$ n/cm$^2$/s (each with an uncertainty of 1.1%). The flux response of our devices is linear over the range tested, and fig. S7 shows the measured signal magnitudes (ΔI) and signal to noise ratio (SNR) four our devices in the access column.

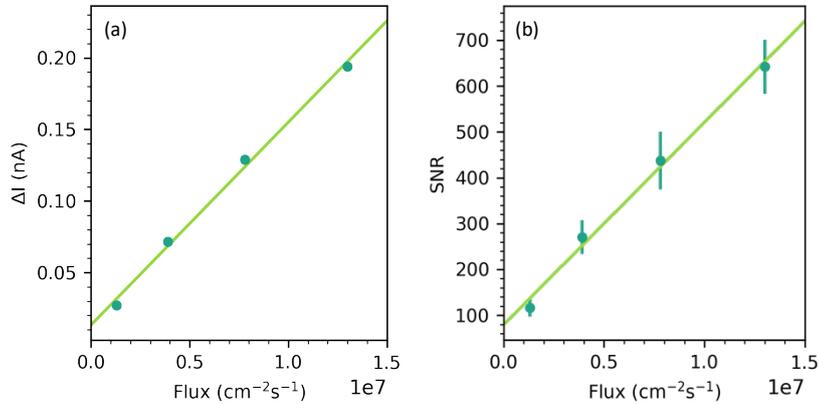

**Figure S7:** a) The signal response and b) the SNR as a function of flux. Both showing linearity from 0.13 to 1.3 ×10$^7$ n/cm$^2$/s.

We observe a 1.5% (4.2%) degradation in the signal (SNR) for our PNDI detectors after an exposure of 4×10$^{10}$ neutrons in the thermal pile. This is equivalent to 5 orders of magnitude higher exposure than the annual atmospheric fluence[9]. Some OSCs have been observed to tolerate large exposures and to exhibit annealing effects[10, 11, 12]. As some applications (for example fission reactor monitoring) will require a detailed understanding of the radiation hardness, this issue merits further detailed study.

The energy spectrum of an AmBe source depends on the amount of material used to house that source, ergo the activity of that source, as discussed in[13]. Here we use 1 and 15 Ci (37 and 555 GBq) AmBe sources with our devices. The lower activity source provides an energy spectrum of predominantly fast neutrons with a small component of lower energy neutrons. Secondary emissions from the larger housing for the 15 Ci source result in an enhanced lower energy neutron component to the spectrum. To make measurements with the source an operator manually manipulated it by placing the source on the detector housing, and subsequently removing the source. This process was repeated to demonstrate the reproducibility of the signal (fig. S8). The source-detector distance was approximately 2.5 cm when the source was placed in situ, and over 2 meters when removed. The source removed neutron fluxes /cm$^2$ are estimated to be less than 74 kBq and 1.1 MBq for the 37 and 555 GBq activities, respectively.

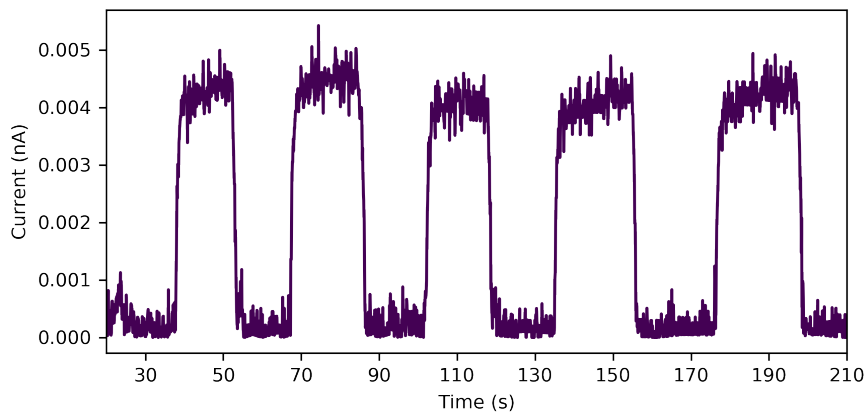

**Figure S8:** Results of exposing a type 1 device to a 555 GBq AmBe source with a -30 V bias applied (with an arbitrary offset). The current pulse increases correspond to the source being placed on the sample box, approximately 2.5 cm from the device under test.

### Variations in neutron flux at NPL

It is important to note that the neutron flux an NPL displayed variations over the timescales of the device exposures. These variations were detected by our devices and do not represent instrumental artefacts. To demonstrate this, we display a typical response from a type 2 device in the thermal pile (fig. S9), showing typical initial "overshoot" as the beam is turned on. The overshoot is confirmed as a real beam current effect independently by the NPL beam monitoring instruments. A feedback system for the accelerator stabilises the beam current within a few seconds for the beamline used for the thermal pile.

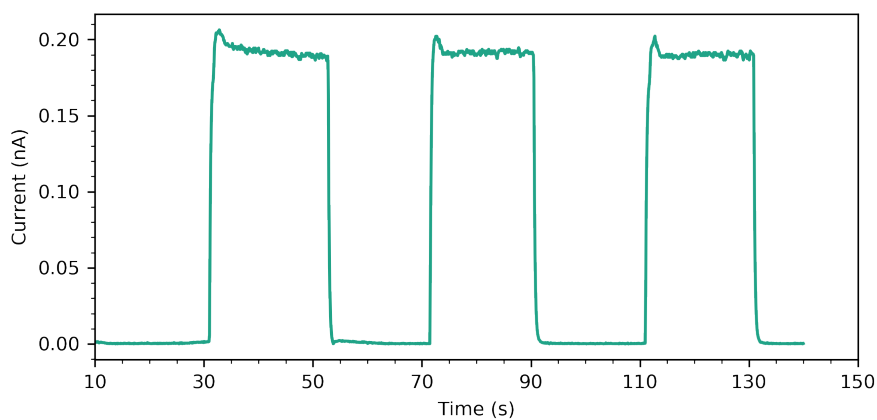

**Figure S9:** Typical type 2 device output from the thermal pile using a -100 V bias. The "overshoot" on switching on the neutron flux can be clearly distinguished from the constant flux operation.

### Device Efficiency Estimation

Using the $^{10}$B thermal neutron cross section, the $^{10}$B average atomic number density and the device thickness we can calculate the probability of thermal neutron capture in the type 2 device as (18±3) %. This value provides an upper limit on any detection efficiency and is referred to by us and other authors as a "quantum efficiency"[14]. By subtracting the type 1 device response from the type 2 device response in the thermal column (fig. 3a) we can measure the signal enhancement due to the $^{10}$B thermal neutron capture. We have previously characterised both types of devices using a known α source and have obtained the gain-efficiency product for α detection as a function of bias (see F. E. Taifakou et. al. for the definition of the product[15]). The signal enhancement due to the $^{10}$B loading can thus be converted to an effective α particle flux by using the average energy of the α particle and $^{7}$Li decay products. Thus, a thermal neutron conversion efficiency can be calculated for the type 2 device. The calculated thermal neutron conversion efficiency is of order of 2 %. This efficiency is considerably lower than the calculated quantum efficiency for the device (which is physically acceptable) and we attribute the difference to inhomogeneous distribution of the $^{10}$B within the type 2 device. There is evidence that the B$_4$C in the type 2 device forms clusters 20 to 100 μm in size (see SI fig. 11). The decay products of thermal neutrons captured and converted within the bulk of such clusters cannot necessarily reach the OSC to deposit their energy and contribute to the current signal, leading to reduced detection efficiency.

Fast neutron detection efficiency estimation requires a detailed simulation of the neutron interactions with the OSC to determine the ratio of energy detected (corresponding to the current obtained) and the energy flux of the beam in a device. For the 0.565keV beam this requires study of H and C elastic scatter, and for 16.5 MeV this requires study of both elastic and inelastic contributions.

### Film characterisation

A Bruker DektakXT profilometer was used to study the roughness of the devices fabricated (fig. S10). This figure shows type 1 and 2 devices, where the colour map output by the profilometer is fixed to the full range (being [−10, +9] for type 1 and [−16, +22] for type 2). The type 2 devices contain B$_4$C, and there is evidence of clusters of this material with scales of tens of μm from the 1 mm × 1 mm surface profile. These are also evident in the corresponding optical microscope images of the OSC film on our device substrates (fig. S11).

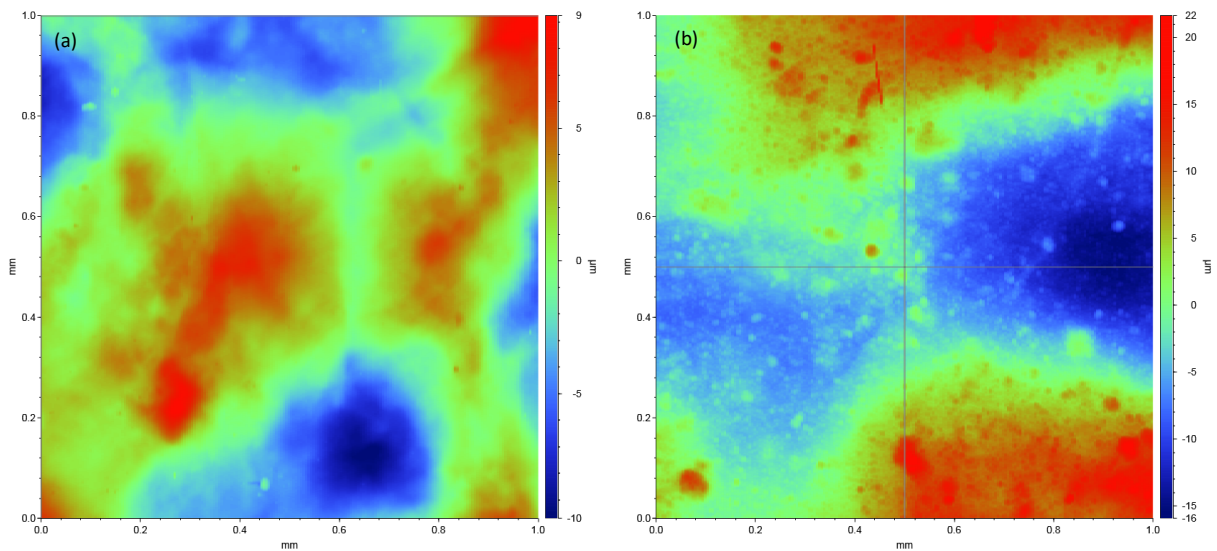

**Figure S10:** Surface profilometry of a) a type 1 device and b) a type 2 device.

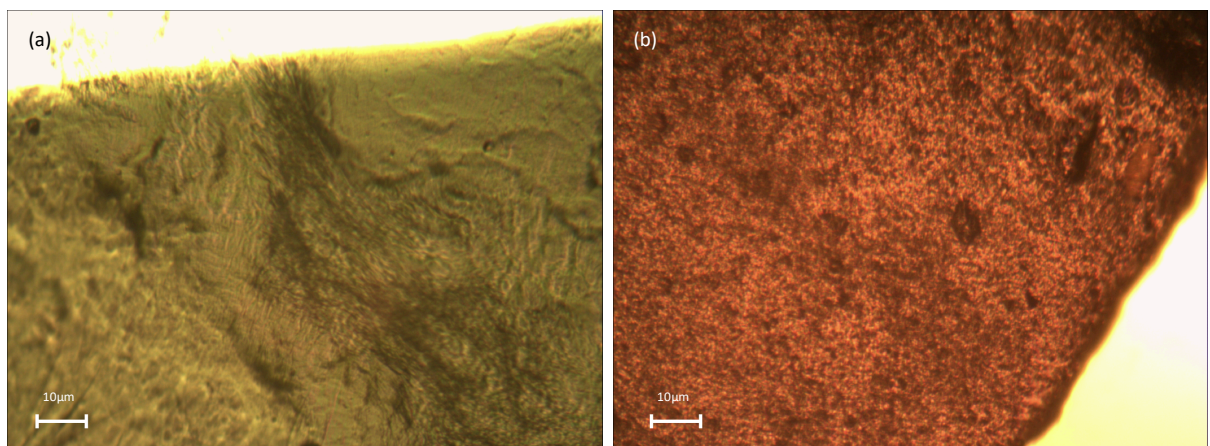

**Figure S11:** Optical microscopy of a) a type 1 device and b) a type 2 device.

References

[1] https://www.ossila.com/products/pndi2hd2t.

[2] L. Gao et al., All-Polymer Solar Cells Based on Absorption-Complementary Polymer Donor and Acceptor with High Power Conversion Efficiency of 8.27%, Adv. Mater., **28**, 1884–1890 (2016); DOI: 10.1002/adma.201504629.

[3] J. Allison et al., Recent developments in GEANT4, Nucl. Instrum. Meth. A **835** (2016) 186.

[4] E. Mendoza, D. Cano-Ott, T. Koi and C. Guerrero, New standard evaluated neutron cross section libraries for the GEANT4 code and first verification, IEEE Trans. Nucl. Science **61** (2014) 2357.

[5] E. Mendoza, D. Cano-Ott, C. Guerrero, and R. Capote, New Evaluated Neutron Cross Section Libraries for the Geant4 Code, IAEA technical report INDC(NDS)-0612, Vienna, 2012. Data available online at Geant4.

[6] E. Mendoza, D. Cano-Ott, Update of the Evaluated Neutron Cross Section Libraries for the Geant4 Code, IAEA technical report INDC(NDS)-0758, Vienna, 2018.

[7] R. A Cecil, B. D. Anderson and R. Madey, Improved predictions of neutron detection efficiency for hydrocarbon scintillators from 1 MeV to about 300 MeV, Nucl. Instrum. Meth. **161** (1979) 439-447.

[8] N. P. Hawkes et al., Additional Characterisation of the thermal neutron pile at the National Physical Laboratory, UK, Radiation Protection and Dosimetry **180** (2018) No 1-4, 25-28.

[9] T. W. Armstrong and B. L. Colborn, Predictions of secondary neutrons and their importance to radiation effects inside the international space station. Radiat. Meas. **33** (2001) 229–234.

[10] G. M. Paterno et al., Neutron Radiation Tolerance of Two Benchmark Thiophene-Based Conjugated Polymers: The Importance of Crystallinity for Organic Avionics, (2017) Jan 23;**7**:41013. doi:10.1038/srep41013.

[11] R. A. Street et al., Radiation induced recombination centers in organic solar cells. Phys. Rev. B **85** (2012) 205211– 205224.

[12] H. N. Raval et al., Investigation of effects of ionizing radiation exposure on material properties of organic semiconducting oligomer – Pentacene, Org. Electron. **14**, (2013) 1467–1476.

[13] J. W. Marsh, D. J. Thomas, M. Burke, High resolution measurements of neutron energy spectra from Am-Be and Am-B neutron sources, Nucl. Instrum. Meth. A **366** (1995) 340-348.

[14] P. Chatzipiroglou, J. L. Keddie, and P. J. Sellin, Boron-Loaded Polymeric Sensor for the Direct Detection of Thermal Neutrons, ACS Applied Materials & Interfaces **12** (29) (2020) 33050-33057.

[15] F. E. Taifakou et al., Solution-Processed Donor–Acceptor Poly(3-hexylthiophene):Phenyl-C61-butyric Acid Methyl Ester Diodes for Low-Voltage α Particle Detection, ACS Appl. Mater. Interfaces, **13**, 5, (2021) 6470–6479.

## Author contributions

Paper Writing: MA, CA, AB, JB, TK, FET, Simulation: MA, CT, FET Fabrication: JB, FET, Lab and Field Tests: AB, JB, TK, FET, CT, Analysis: AB, JB, TK, FET, Interpretation: AB, FET, TK, Project oversight: AB, TK

## Data access statement

All relevant data are provided with the paper and its SI.